\documentclass[preprint,aps,prd,USpaper,nofootinbib,showpacs,superscriptaddress]{revtex4-2}
\usepackage{amsmath,amssymb,graphicx,mathrsfs,xcolor}
\usepackage{bm} 
\usepackage{feynmp}

\usepackage{ifpdf}
\ifpdf
\fi
\makeatletter
\def\endfmffile{%
  \fmfcmd{\p@rcent\space the end.^^J%
          end.^^J%
          endinput;}%
  \if@fmfio
    \immediate\closeout\@outfmf
  \fi
  \IfFileExists{\thefmffile.mp}{\immediate\write18{mpost \thefmffile}}{}
  \let\thefmffile\relax
}
\makeatother
\newcommand{\nn}{\nonumber\\}

\newcommand{\Vg}{V_{\mbox{\scriptsize gluon}} }
\newcommand{\Vc}{V_{\mbox{\scriptsize conf}}}

\newcommand{\ben}{\begin{displaymath}}
\newcommand{\een}{\end{displaymath}}
\newcommand{\be}{\begin{equation}}
\newcommand{\ee}{\end{equation}}
\newcommand{\bea}{\begin{eqnarray}}
\newcommand{\eea}{\end{eqnarray}}

\newcommand{\Kf}{K}

\newcommand{\PPsi}{{\it\Psi}}

\newcommand{\TT}{{\cal T}}
\newcommand{\KK}{{\cal K}}

\newcommand{\bc}{\begin{center}}
\newcommand{\ec}{\end{center}}

\newcommand{\eqn}[1]{\label{#1}}
\newcommand{\eq}[1]{Eq.~(\ref{#1})}
\newcommand{\eqs}[1]{Eqs.~(\ref{#1})}
\newcommand{\fign}[1]{\label{#1}}
\newcommand{\fig}[1]{Fig.~\ref{#1}}

\begin{document}
\title{Unified triquark equations}
\author{A. N. Kvinikhidze}
\email{sasha\_kvinikhidze@hotmail.com}
\affiliation{Andrea Razmadze Mathematical Institute of Tbilisi State University, 6, Tamarashvili Str., 0186 Tbilisi, Georgia}
\affiliation{
College of Science and Engineering,
 Flinders University, Bedford Park, SA 5042, Australia}
\author{B. Blankleider}
\email{boris.blankleider@flinders.edu.au}
\affiliation{
College of Science and Engineering,
 Flinders University, Bedford Park, SA 5042, Australia}

\date{\today}

\begin{abstract}
We derive covariant equations describing the three-quark bound state in terms of quark and diquark degrees of freedom. The equations are exact in the approximation where three-body forces are neglected. A feature of these equations is that they unify two often-used but seemingly unrelated approaches that model baryons as quark-diquark systems; namely, (i) the approach using Poincar\'{e} covariant quark+diquark Faddeev equations driven by a one-quark-exchange kernel [pioneered by Cahill {\it et al.},  Austral.\ J.\ Phys.\ {\bf 42}, 129 (1989) and Reinhardt, Phys.\ Lett.\ B {\bf 244}, 316 (1990)],  and  (ii) the approach using the quasipotential quark-diquark bound-state equation where the kernel consists of the lowest-order contribution from an underlying quark-quark potential [pioneered by Ebert {\it et al.}, Z.\ Phys.\ C {\bf 76} 111 (1997)]. In particular, we show that each of these approaches corresponds to the unified equations with its kernel taken in different, non-overlapping, approximations.
\end{abstract}

\maketitle
\newpage

\section{Introduction}
The use of diquarks as effective degrees of freedom in describing hadrons has a long history, as evidenced by a number of reviews over the last thirty years \cite{Anselmino:1992vg, Alkofer:2000wg,Eichmann:2016yit, Barabanov:2020jvn}. Documented are different quark-diquark approaches for baryons, but to the best of our knowledge, no attempt has been made for their comparison on the basis  of quantum field theory. In the present work, we would like to make such a comparison, demonstrating that two of the most often-used quark-diquark models of baryons, which have usually been considered as separate, unrelated models of baryons, are in fact two non-overlapping parts of the same quark-diquark model.
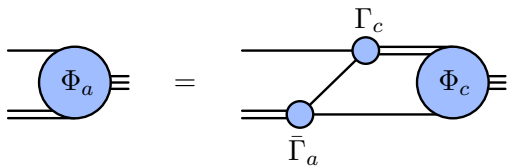
\begin{figure}[b]
\begin{center}
\begin{fmffile}{cahill}
\[
\parbox{16mm}{
\begin{fmfgraph*}(16,8.5)
\fmfstraight
\fmfleftn{l}{3}\fmfrightn{r}{3}\fmfbottomn{b}{3}\fmftopn{t}{3}
\fmf{phantom}{l1,m1,r1}
\fmf{phantom}{l2,m2,r2}
\fmf{phantom}{l3,m3,r3}
\fmffreeze
\fmfshift{2.5 right}{m1}
\fmfshift{2.5 right}{m2}
\fmfshift{2.5 right}{m3}
\fmfi{plain}{vpath (__m1,__l1) shifted (thick*(0,.7))}
\fmfi{plain}{vpath (__m1,__l1) shifted (thick*(0,-.7))}
\fmf{plain}{l3,m3}
\fmfv{d.sh=circle,d.f=empty,d.si=27.,label=$\vspace{0mm}\hspace{-4mm}\Phi_a$,background=(.6235,,.7412,,1)}{m2}
\fmfi{plain}{vpath (__m2,__r2) shifted (thick*(0,1.2))}
\fmfi{plain}{vpath (__m2,__r2) shifted (thick*(0,0.))}
\fmfi{plain}{vpath (__m2,__r2) shifted (thick*(0,-1.2))}
\end{fmfgraph*}}
\hspace{5mm} = \hspace{5mm}
\parbox{35mm}{
\begin{fmfgraph*}(35,8.5)
\fmfstraight
\fmfleftn{l}{3}\fmfrightn{r}{3}\fmfbottomn{b}{5}\fmftopn{t}{5}
\fmf{phantom}{l1,v1,w1,m1,r1}
\fmf{phantom}{l2,v2,w2,m2,r2}
\fmf{phantom}{l3,v3,w3,m3,r3}
\fmffreeze
\fmfshift{3 left}{v1}
\fmfshift{3 left}{w3}
\fmfshift{5 right}{m1}
\fmfshift{5 right}{m2}
\fmfshift{5 right}{m3}
\fmf{plain}{v1,w3}
\fmf{plain}{v1,m1}
\fmfi{plain}{vpath (__v1,__l1) shifted (thick*(0,.7))}
\fmfi{plain}{vpath (__v1,__l1) shifted (thick*(0,-.7))}
\fmf{plain}{l3,w3}
\fmfi{plain}{vpath (__w3,__m3) shifted (thick*(0,.7))}
\fmfi{plain}{vpath (__w3,__m3) shifted (thick*(0,-.7))}
\fmfv{d.sh=circle,d.f=empty,d.si=10,label=$\bar \Gamma_a$,l.a=-80.,l.d=3mm,background=(.6235,,.7412,,1)}{v1}
\fmfv{d.sh=circle,d.f=empty,d.si=10,label=$\Gamma_c$,l.a=80.,l.d=2.5mm,background=(.6235,,.7412,,1)}{w3}
\fmfv{d.sh=circle,d.f=empty,d.si=27.,label=$\vspace{0mm}\hspace{-4mm}\Phi_c$,background=(.6235,,.7412,,1)}{m2}
\fmfi{plain}{vpath (__m2,__r2) shifted (thick*(0,1.2))}
\fmfi{plain}{vpath (__m2,__r2) shifted (thick*(0,0.))}
\fmfi{plain}{vpath (__m2,__r2) shifted (thick*(0,-1.2))}
\end{fmfgraph*}}
\]
\end{fmffile}   
\vspace{-3mm}

\caption{\fign{Reg}  Poincar\'{e} covariant quark+diquark Faddeev equations of Ref.\ \cite{Cahill:1988dx, Reinhardt:1989rw}.  The amplitudes $\Phi_a$ and $\Phi_c$ are Faddeev components coupling the baryon to quark (single line) and diquark (double-line) states. The equation kernel corresponds to one-quark-exchange, with $\Gamma_c$ and $\bar\Gamma_a$ being vertex functions describing the disinegration and formation of the diquark.}
\end{center}
\end{figure}

The first of these models, proposed more than thirty years ago  \cite{Cahill:1988dx, Reinhardt:1989rw}, is based on a description of three quarks using covariant Faddeev equations where the quark-quark t matrix is approximated by one or more diquark-pole terms (i.e.,  terms with a pole at the diquark mass, and with a residue that is expressed as an outer product of form factors $\Gamma$ and $\bar\Gamma$ for the transition between the diquark and two free quarks). The resulting coupled set of bound-state equations are illustrated in \fig{Reg}. Sometimes referred to as Poincar\'{e} covariant quark+diquark Faddeev equations \cite{Chen:2012qr,Barabanov:2020jvn}, and sometimes as quark-diquark Bethe-Salpeter equations \cite{Oettel:2001kd,Eichmann:2016hgl}, they have been used extensively over the years, see \cite{Oettel:1998bk,Ahlig:2000qu,Oettel:2001kd,Eichmann:2007nn,Nicmorus:2008vb,Wilson:2011aa,Chen:2012qr,Wang:2013wk,Segovia:2015ufa,Eichmann:2016hgl,Chen:2019fzn,Liu:2022nku} for a representative selection of works.

The second model, proposed more that 25 years ago \cite{Ebert:1996ec}, is a relativistic description of the quark-diquark system using  quasipotential equations (we will refer to it as the ``quasipotential quark-diquark model"), which has likewise been often used over the years \cite{Ebert:1996ec,Ebert:2004ck,Ebert:2005ip,Ebert:2005xj,Ebert:2007nw,Ebert:2008kb,Ebert:2009zza,Faustov:2015eba,Faustov:2018vgl,Faustov:2020gun,Faustov:2021qqf}. In this model one first constructs a quark-quark potential of the form
\be
V_{qq} = V_{\mbox{\scriptsize gluon}} + V_{\mbox{\scriptsize conf}}   \eqn{t_moscow}
\ee
where $\Vg$ is the quark-quark ($qq$) one-gluon-exchange potential and $\Vc$ is a local confining  potential, and then uses this to construct the quark-diquark potential which then forms the kernel of a relativistic quark-diquark quasipotential equation for the baryon. Illustrated in \fig{faust}, this bound-state equation again has the form of a Faddeev equation, but with a kernel corresponding to a single rescattering of two quarks via potential $V_{qq}$  (specified in the diagram as quarks a and c scattering via a potential $K_b$ ($= V_{qq}$) with quark $b$ being a spectator).

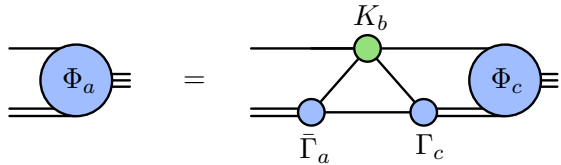
\begin{figure}[t]
\begin{center}
\begin{fmffile}{faust}
\[
\parbox{16mm}{
\begin{fmfgraph*}(16,8.5)
\fmfstraight
\fmfleftn{l}{3}\fmfrightn{r}{3}\fmfbottomn{b}{3}\fmftopn{t}{3}
\fmf{phantom}{l1,m1,r1}
\fmf{phantom}{l2,m2,r2}
\fmf{phantom}{l3,m3,r3}
\fmffreeze
\fmfshift{2.5 right}{m1}
\fmfshift{2.5 right}{m2}
\fmfshift{2.5 right}{m3}
\fmfi{plain}{vpath (__m1,__l1) shifted (thick*(0,.7))}
\fmfi{plain}{vpath (__m1,__l1) shifted (thick*(0,-.7))}
\fmf{plain}{l3,m3}
\fmfv{d.sh=circle,d.f=empty,d.si=27.,label=$\vspace{0mm}\hspace{-4mm}\Phi_a$,background=(.6235,,.7412,,1)}{m2}
\fmfi{plain}{vpath (__m2,__r2) shifted (thick*(0,1.2))}
\fmfi{plain}{vpath (__m2,__r2) shifted (thick*(0,0.))}
\fmfi{plain}{vpath (__m2,__r2) shifted (thick*(0,-1.2))}
\end{fmfgraph*}}
\hspace{6mm} = \hspace{5mm} 
\parbox{40mm}{
\begin{fmfgraph*}(40,8.5)
\fmfstraight
\fmfleftn{l}{3}\fmfrightn{r}{3}\fmfbottomn{b}{6}\fmftopn{t}{6}
\fmf{phantom}{l1,v1,w1,x1,m1,r1}
\fmf{phantom}{l2,v2,w2,x2,m2,r2}
\fmf{phantom}{l3,v3,w3,x3,m3,r3}
\fmffreeze
\fmfshift{1.5 left}{w3}
\fmfshift{3 left}{x1}
\fmfshift{5 right}{m1}
\fmfshift{5 right}{m2}
\fmfshift{5 right}{m3}
\fmf{plain}{v1,w3,x1}
\fmf{plain}{v3,m3}
\fmfi{plain}{vpath (__v1,__l1) shifted (thick*(0,.7))}
\fmfi{plain}{vpath (__v1,__l1) shifted (thick*(0,-.7))}
\fmf{plain}{l3,w3}
\fmf{plain}{v1,x1}
\fmfi{plain}{vpath (__x1,__m1) shifted (thick*(0,.7))}
\fmfi{plain}{vpath (__x1,__m1) shifted (thick*(0,-.7))}
\fmfv{d.sh=circle,d.f=empty,d.si=10,label=$\bar\Gamma_a$,l.a=-80.,l.d=3.mm,background=(.6235,,.7412,,1)}{v1}
\fmfv{d.sh=circle,d.f=empty,d.si=10,label=$\Gamma_c$,l.a=-74.,l.d=3.1mm,background=(.6235,,.7412,,1)}{x1}
\fmfv{d.sh=circle,d.f=empty,d.si=10,label=$K_b$,l.a=80.,l.d=2.5mm,background=(0.580392,, 0.878431,, 0.490196)}{w3}
\fmfv{d.sh=circle,d.f=empty,d.si=27.,label=$\vspace{0mm}\hspace{-4mm}\Phi_c$,background=(.6235,,.7412,,1)}{m2}
\fmfshift{2 right}{r2}
\fmfi{plain}{vpath (__m2,__r2) shifted (thick*(0,1.2))}
\fmfi{plain}{vpath (__m2,__r2) shifted (thick*(0,0.))}
\fmfi{plain}{vpath (__m2,__r2) shifted (thick*(0,-1.2))}
\end{fmfgraph*}}
\]
\end{fmffile}   
\vspace{-3mm}

\caption{\fign{faust} Equations corresponding to the quasipotential quark-diquark model of Ref.\ \cite{Ebert:1996ec}.  Similar to  \fig{Reg}, amplitudes $\Phi_a$ and $\Phi_c$ are Faddeev components coupling the baryon to quark-diquark states, with $\Gamma_c$ and $\bar\Gamma_a$ being diquark vertex functions. However, the kernel of this equation involves a single scattering of two quarks (quarks a and c in this case) via a potential $K_b$.}
\end{center}
\end{figure}
In the following, we derive covariant triquark bound-state equations that are exact for the case where three-body forces are neglected. These equations are illustrated in \fig{kvin}, and have the form of Faddeev equations where the kernel consists of an infinite series involving successive numbers of quark-exchanges between quark-diquark states. It is evident that the  Poincar\'{e} covariant quark+diquark Faddeev equations of Ref.\ \cite{Cahill:1988dx, Reinhardt:1989rw} correspond to keeping just the first term in the infinite series, and the quasipotential equations of Ref.\ \cite{Ebert:1996ec} correspond to keeping just the second term in this series. As such, our triquark equations unify these two popular approaches for modeling baryons in terms of quark and diquark degrees of freedom. Moreover, it is evident that these two approaches should not be viewed as unrelated competing models of baryons, but rather, as different approximations of the same model. Indeed, any competition between these models at describing data, needs to be assessed by comparing their kernels, as these are non-overlapping terms appearing in the unified equations.

Ideally, the two approaches should be combined, with a kernel that is the sum of the first two terms of the infinite series illustrated in \fig{kvin}. Additionally, in light of the unification embodied in \fig{kvin}, all sorts of form-factors (electromagnetic, axial-vector, pseudoscalar, etc.) should also be unified correspondingly. This can be done by gauging the equation of \fig{kvin}  \cite{Kvinikhidze:1998xn,*Kvinikhidze:1999xp}, thereby obtaining contributions to the baryon form factors coming from both of the first two kernels in this figure. By contrast, the current situation is that the baryon form factors are being pursued intensively in each of the two approaches separately (just in the last few years, the Poincar\'{e} covariant quark+diquark Faddeev equations have been used to calculate such form factors in Refs.\ \cite{Chen:2018nsg,Lu:2019bjs,Cui:2020rmu,Chen:2020wuq,Raya:2021pyr,Chen:2021guo,Chen:2022odn,Yin:2023kom}
and the quasipotential quark-diquark approach has been used to calculate them in Refs.\ \cite{Faustov:2016yza,Faustov:2017ous,Faustov:2017wbh,Faustov:2018ahb, Faustov:2019ddj,Faustov:2020thr,Galkin:2020mjm,Davydov:2022glx}). 

It is worth noting that analogous unified equations were derived for the tetraquark \cite{Kvinikhidze:2023djp}.

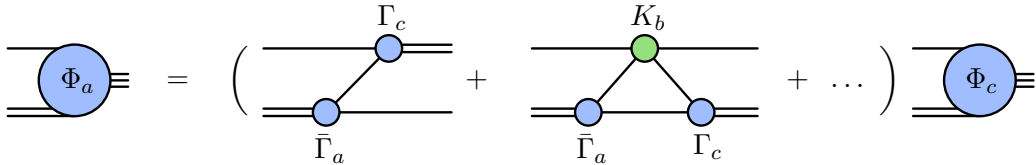
\begin{figure}[t]
\begin{center}
\begin{fmffile}{kb}
\[
\parbox{16mm}{
\begin{fmfgraph*}(16,8.5)
\fmfstraight
\fmfleftn{l}{3}\fmfrightn{r}{3}\fmfbottomn{b}{3}\fmftopn{t}{3}
\fmf{phantom}{l1,m1,r1}
\fmf{phantom}{l2,m2,r2}
\fmf{phantom}{l3,m3,r3}
\fmffreeze
\fmfshift{2.5 right}{m1}
\fmfshift{2.5 right}{m2}
\fmfshift{2.5 right}{m3}
\fmfi{plain}{vpath (__m1,__l1) shifted (thick*(0,.7))}
\fmfi{plain}{vpath (__m1,__l1) shifted (thick*(0,-.7))}
\fmf{plain}{l3,m3}
\fmfv{d.sh=circle,d.f=empty,d.si=27.,label=$\vspace{0mm}\hspace{-4mm}\Phi_a$,background=(.6235,,.7412,,1)}{m2}
\fmfi{plain}{vpath (__m2,__r2) shifted (thick*(0,1.2))}
\fmfi{plain}{vpath (__m2,__r2) shifted (thick*(0,0.))}
\fmfi{plain}{vpath (__m2,__r2) shifted (thick*(0,-1.2))}
\end{fmfgraph*}}
\hspace{4mm} = \hspace{4mm}\left(\hspace{2mm} 
\parbox{25mm}{
\begin{fmfgraph*}(25,8.5)
\fmfstraight
\fmfleftn{l}{3}\fmfrightn{r}{3}\fmfbottomn{b}{5}\fmftopn{t}{5}
\fmf{phantom}{l1,v1,w1,r1}
\fmf{phantom}{l2,v2,w2,r2}
\fmf{phantom}{l3,v3,w3,r3}
\fmffreeze
\fmfshift{0 left}{v1}
\fmfshift{0 right}{w3}
\fmf{plain}{v1,w3}
\fmfi{plain}{vpath (__v1,__l1) shifted (thick*(0,.7))}
\fmfi{plain}{vpath (__v1,__l1) shifted (thick*(0,-.7))}
\fmf{plain}{l3,w3}
\fmf{plain}{r1,v1}
\fmfi{plain}{vpath (__w3,__r3) shifted (thick*(0,.7))}
\fmfi{plain}{vpath (__w3,__r3) shifted (thick*(0,-.7))}
\fmfv{d.sh=circle,d.f=empty,d.si=10,label=$\bar \Gamma_a$,l.a=-80.,l.d=3mm,background=(.6235,,.7412,,1)}{v1}
\fmfv{d.sh=circle,d.f=empty,d.si=10,label=$\Gamma_c$,l.a=80.,l.d=2.5mm,background=(.6235,,.7412,,1)}{w3}
\end{fmfgraph*}}
\hspace{1mm} + \hspace{5mm} 
\parbox{30mm}{
\begin{fmfgraph*}(30,8.5)
\fmfstraight
\fmfleftn{l}{3}\fmfrightn{r}{3}\fmfbottomn{b}{6}\fmftopn{t}{6}
\fmf{phantom}{l1,v1,x1,w1,r1}
\fmf{phantom}{l2,v2,x2,w2,r2}
\fmf{phantom}{l3,v3,x3,w3,r3}
\fmffreeze
\fmfshift{0 left}{x1}
\fmf{plain}{v1,x3,w1}
\fmf{plain}{x3,r3}
\fmfi{plain}{vpath (__v1,__l1) shifted (thick*(0,.7))}
\fmfi{plain}{vpath (__v1,__l1) shifted (thick*(0,-.7))}
\fmf{plain}{l3,x3}
\fmf{plain}{v1,w1}
\fmfi{plain}{vpath (__w1,__r1) shifted (thick*(0,.7))}
\fmfi{plain}{vpath (__w1,__r1) shifted (thick*(0,-.7))}
\fmfv{d.sh=circle,d.f=empty,d.si=10,label=$\bar\Gamma_a$,l.a=-80.,l.d=3.mm,background=(.6235,,.7412,,1)}{v1}
\fmfv{d.sh=circle,d.f=empty,d.si=10,label=$\Gamma_c$,l.a=-74.,l.d=3.1mm,background=(.6235,,.7412,,1)}{w1}
\fmfv{d.sh=circle,d.f=empty,d.si=10,label=$K_b$,l.a=80.,l.d=2.5mm,background=(0.580392,, 0.878431,, 0.490196)}{x3}
\end{fmfgraph*}}
\hspace{3mm} + \hspace{2mm} \dots\hspace{2mm} \right)\hspace{1mm}
\parbox{16mm}{
\begin{fmfgraph*}(16,8.5)
\fmfstraight
\fmfleftn{l}{3}\fmfrightn{r}{3}\fmfbottomn{b}{3}\fmftopn{t}{3}
\fmf{phantom}{l1,m1,r1}
\fmf{phantom}{l2,m2,r2}
\fmf{phantom}{l3,m3,r3}
\fmffreeze
\fmfshift{2.5 right}{m1}
\fmfshift{2.5 right}{m2}
\fmfshift{2.5 right}{m3}
\fmfi{plain}{vpath (__m1,__l1) shifted (thick*(0,.7))}
\fmfi{plain}{vpath (__m1,__l1) shifted (thick*(0,-.7))}
\fmf{plain}{l3,m3}
\fmfv{d.sh=circle,d.f=empty,d.si=27.,label=$\vspace{0mm}\hspace{-4mm}\Phi_c$,background=(.6235,,.7412,,1)}{m2}
\fmfi{plain}{vpath (__m2,__r2) shifted (thick*(0,1.2))}
\fmfi{plain}{vpath (__m2,__r2) shifted (thick*(0,0.))}
\fmfi{plain}{vpath (__m2,__r2) shifted (thick*(0,-1.2))}
\end{fmfgraph*}}
\]
\end{fmffile}   
\vspace{-3mm}

\caption{\fign{kvin}  Unified quark+diquark equations derived in this paper. The kernel of this equation is an infinite series whose first two terms, separately,  correspond to the model of  Ref.\ \cite{Cahill:1988dx, Reinhardt:1989rw} as illustrated in \fig{Reg}, and the model of Ref.\ \cite{Ebert:1996ec} as illustrated in \fig{faust}, respectively.}
\end{center}
\end{figure}

\section{Derivation}

\subsection{Triquark equations for distinguishable quarks}
For clarity of presentation, we first consider the case of three distinguishable quarks. To describe such a system where only pairwise interactions are taken into account, we follow the formulation of Faddeev \cite{Faddeev:1960su}. Thus,
assigning labels 1, 2 and 3 to the quarks, and using a notation where $(abc)$ is a cyclic permutation of $(123)$, the three-body ($3q$) kernel, $K$, is written as
\be
K=\sum_{a} K_{a}    \eqn{pair}
\ee
where $K_a$ is the kernel where quarks $b$ and $c$ are interacting while quark $a$ is a spectator, as illustrated in \fig{K}.

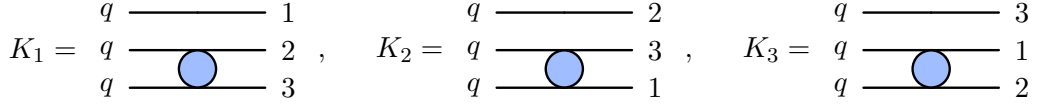
\begin{figure}[t]
\begin{center}
\begin{fmffile}{K}
\begin{align*}
\Kf_1&=\hspace{5mm}
\parbox{20mm}{
\begin{fmfgraph*}(18,10)
\fmfstraight
\fmfleft{f3,f2,f1}\fmfright{i3,i2,i1}
\fmf{plain,tension=1.3}{i1,v1,f1}
\fmf{plain,tension=1.3}{i2,v2,f2}
\fmf{plain,tension=1.3}{i3,v3,f3}
\fmffreeze
\fmf{phantom,tension=1.3}{v2,v23,v3}
\fmfv{d.s=circle,d.f=empty,d.si=14,background=(.6235,,.7412,,1)}{v23}
\fmfv{label=$1$,l.a=0}{i1}
\fmfv{label=$2$,l.a=0}{i2}
\fmfv{label=$3$,l.a=0}{i3}
\fmfv{label=$q$,l.a=180}{f1}
\fmfv{label=$q$,l.a=180}{f2}
\fmfv{label=$q$,l.a=180}{f3}
\end{fmfgraph*}}
\hspace{6mm},
\hspace{6mm}
\Kf_2=\hspace{5mm}
\parbox{20mm}{
\begin{fmfgraph*}(18,10)
\fmfstraight
\fmfleft{f3,f2,f1}\fmfright{i3,i2,i1}
\fmf{plain,tension=1.3}{i1,v1,f1}
\fmf{plain,tension=1.3}{i2,v2,f2}
\fmf{plain,tension=1.3}{i3,v3,f3}
\fmffreeze
\fmf{phantom,tension=1.3}{v2,v23,v3}
\fmfv{d.s=circle,d.f=empty,d.si=14,background=(.6235,,.7412,,1)}{v23}
\fmfv{label=$2$,l.a=0}{i1}
\fmfv{label=$3$,l.a=0}{i2}
\fmfv{label=$1$,l.a=0}{i3}
\fmfv{label=$q$,l.a=180}{f1}
\fmfv{label=$q$,l.a=180}{f2}
\fmfv{label=$q$,l.a=180}{f3}
\end{fmfgraph*}}
\hspace{6mm}, \hspace{6mm}
\Kf_3=\hspace{5mm}
\parbox{20mm}{
\begin{fmfgraph*}(18,10)
\fmfstraight
\fmfleft{f3,f2,f1}\fmfright{i3,i2,i1}
\fmf{plain,tension=1.3}{i1,v1,f1}
\fmf{plain,tension=1.3}{i2,v2,f2}
\fmf{plain,tension=1.3}{i3,v3,f3}
\fmffreeze
\fmf{phantom,tension=1.3}{v2,v23,v3}
\fmfv{d.s=circle,d.f=empty,d.si=14,background=(.6235,,.7412,,1)}{v23}
\fmfv{label=$3$,l.a=0}{i1}
\fmfv{label=$1$,l.a=0}{i2}
\fmfv{label=$2$,l.a=0}{i3}
\fmfv{label=$q$,l.a=180}{f1}
\fmfv{label=$q$,l.a=180}{f2}
\fmfv{label=$q$,l.a=180}{f3}
\end{fmfgraph*}}
\end{align*}
\end{fmffile}   
\vspace{-3mm}

\caption{\fign{K}  Structure of the terms $\Kf_a$ ($a=1, 2, 3$) making up the three-body kernel $K$ where only two-body forces are included. The coloured circles represent two-body kernels $K_{bc}$ for the scattering of quarks $b$ and $c$, as indicated.}
\end{center}
\end{figure}
 
The $3q$  bound-state wave function for distinguishable quarks is then
\be
 \Psi= G_0 K \Psi       \eqn{dist2q2q}
 \ee
where $G_0$ is the fully disconnected part of the full $3q$ Green function $G$.
The three-body kernels $K_{a}$ can be used to define the Faddeev components $\Psi_a$ as
\be
\Psi_a = G_0 K_a  \Psi,
\ee
so that
\be
\Psi = \sum_a \Psi_a.
\ee
From \eq{dist2q2q} follow Faddeev's equations for the components,
\be
\Psi_a=\sum_b G_0 T_a \bar\delta_{ab}\Psi_b    \eqn{PsidFad}
\ee
where $\bar\delta_{ab}=1-\delta_{ab}$ and $T_a$ is the t matrix corresponding to kernel $\Kf_a$, so that
\be
 T_a = K_a + K_a G_0 T_a.      \eqn{BSE_Tid}
\ee

Assuming that the $qq$ interaction admits the creation of a diquark, the Green function $G_a$ describing the scattering of quarks $b$ and $c$, will contain a corresponding pole at the diquark mass, so that one can write
\be
G_a = G_a^P + G_a^R
\ee
where $G_a^P$ is the Green function's pole term while $G_a^R$ is its regular part. Then, because
\be
 T_a = K_a + K_a G_a K_a,      \eqn{BSE_Kid}
\ee
the t matrix $T_a$ can be written as
\be
T_a = K_a + T_a^P + T_a^C    \eqn{Tdecomp}
\ee 
where $T_a^P$ is $T_a$'s pole term, while the sum $K_a + T_a^C$ constitutes its regular part. It is important to note that there is no overcounting in this decomposition; that is,  the terms $K_a$, $T^P_a$ and $T^C_a$ do not overlap.  Note that in the case of unconfined quarks, the analytic structure of $T_a$ would be represented by its pole part, $T_a^P$, its part with the 2$q$ branch point, $T_a^C$, and the part $K_a$ again with a branch point, but above the $2q$ mass.

We write \eq{PsidFad} in matrix form as
\be
\PPsi = \TT \PPsi     \eqn{PPsi}
\ee
where $\PPsi$ is a column matrix of elements $\Psi_a$, and $\TT$ is a square matrix whose $(a,b)$'th element is $\TT_{ab} =  G_0 T_a \bar\delta_{ab}$. Similarly we write \eq{Tdecomp} in matrix form as
\be
\TT= \KK + \TT^P + \TT^C    \eqn{TTdecomp}
\ee
where $\KK_{ab} =  G_0 K_a \bar\delta_{ab}$, $\TT^P_{ab} =  G_0 T_a^P \bar\delta_{ab}$, and $\TT^C_{ab} =  G_0 T_a{}^C \bar\delta_{ab}$.
Equation (\ref{PPsi}) can then be recast as
\be
\PPsi = (1-\KK -\TT^C)^{-1} \TT^P \PPsi .  \eqn{recast}
\ee

 Using the separable form of the pole term,
 \be
 T_a^P= \Gamma_a D_a \bar\Gamma_a
 \ee
where $\Gamma_a$ (similarly $\bar\Gamma_a$) and $D_a$ are the diquark form factor and propagator, respectively, \eq{recast} implies that
\be
\Phi_a =\sum_{bc}\bar\Gamma_a \bar\delta_{ab} \left[(1-\KK -\TT^C)^{-1}\right]_{bc} G_0 \Gamma_c D_c  \Phi_c  \eqn{3q-F+G}
\ee
where
\be
\Phi_a = \sum_b \bar\Gamma_a \bar\delta_{ab} {\Psi}_b.
\ee
Expanding the inverse term in \eq{3q-F+G} as
\be
(1-\KK -\TT^C)^{-1} = 1+\KK +\TT^C + \dots,
\ee
we obtain
\begin{align}
\Phi_a &=\sum_{bc}\bar\Gamma_a \bar\delta_{ab} (\delta_{bc}+G_0K_b \bar\delta_{bc}+\dots ) G_0 \Gamma_c D_c  \Phi_c,
\eqn{unified}
\end{align}
which is illustrated in \fig{kvin}.

It is apparent that the first two terms of this series correspond to the models of Refs.\ \cite{Cahill:1988dx, Reinhardt:1989rw} and  Ref.\ \cite{Ebert:1996ec}, respectively. Indeed, keeping just the first term in the series results in the bound-state equation
\begin{align}
\Phi_a &=\sum_{b}\bar\Gamma_a \bar\delta_{ab} G_0 \Gamma_b D_b  \Phi_b
\end{align}
which is illustrated in \fig{Reg} and coincides with the Poincar\'{e} covariant quark+diquark Faddeev equations of Ref.\ \cite{Cahill:1988dx, Reinhardt:1989rw}, and keeping just the second term in the series results in the bound-state equation
\begin{align}
\Phi_a &=\sum_{bc}\bar\Gamma_a \bar\delta_{ab} G_0K_b \bar\delta_{bc} G_0 \Gamma_c D_c  \Phi_c,
\end{align}
which is illustrated in \fig{faust} and coincides with the quasipotential quark-diquark equations of Ref.\ \cite{Ebert:1996ec}.

Although each of the approaches of Refs.\ \cite{Cahill:1988dx, Reinhardt:1989rw} and  Ref.\ \cite{Ebert:1996ec}, can be viewed as different approximations of the same unified equations, \eq{unified}, the reality is that the quark-diquark picture of a baryon is described by a kernel that consists of at least the sum of the first two terms of the series in \eq{unified}. This observation should clarify the true picture of quark-diquark dynamics in baryons.

\subsection{Triquark equations for indistinguishable quarks}

To take into account the antisymmetry of identical quarks, we first note that the Faddeev equations for distinguishable particles,   \eq{PsidFad}, possess fully antisymmetric solutions (as well as symmetric ones) where the component wave functions have the symmetry properties
\begin{alignat}{3}
P_{23} \Psi_1 &= - \Psi_1,  &\hspace{8mm}  P_{12} \Psi_1 &= - \Psi_2, & \hspace{8mm} P_{31} \Psi_1 &= - \Psi_3, \nn
P_{31} \Psi_2 &= - \Psi_2,   & P_{23} \Psi_2 &= - \Psi_3,  &  P_{12} \Psi_2 &= - \Psi_1, \nn
P_{12} \Psi_3 &= - \Psi_3,   & P_{31} \Psi_3 &= - \Psi_1,  & P_{23} \Psi_3 &= - \Psi_2,   \eqn{sym}
\end{alignat}
where $P_{ab}$  is the operator that exchanges the quantum numbers of particles $a$ and $b$. Choosing a solution  with these symmetry properties, \eq{PsidFad} for $\Psi_1$ reduces to
\begin{align}
\Psi_1 &= -G_0 T_1  P_{12}\Psi_1   \eqn{Psi1}
\end{align}
where $T_1$ results from antisymmetrizing the t matrix for distinguishable particles, $T_1^d$, using
\be
T_1=(1-P_{23}) T_1^d.
\ee
Equation (\ref{Psi1}) can be seen most easily by using \eqs{sym}:
\begin{align}
\Psi_1 &= G_0 T_1^d (\Psi_2 + \Psi_3)  = G_0 T_1^d  (1-P_{23})\Psi_2 \nn
&= G_0 (1-P_{23}) T_1^d  \Psi_2 \nn
&= -G_0 (1-P_{23}) T_1^d  P_{12}\Psi_1.
\end{align}

We can then again express $T_1$ as 
\be
T_1=K_1+T^P_1+T^C_1,
\ee
where $T^P_1$ and $K_1+T^C_1$ are the pole and regular parts of $T_1$, but this time with all quantities antisymmetric under the interchange of quark 2 and 3's quantum numbers. 
Equation (\ref{Psi1}) can then be recast as
 \be
 \Psi_1= - \left[1+(\KK_1+\TT^C_1) P_{12}\right]^{-1}\TT^P_1P_{12}\Psi_1
 \ee
 where $\KK_1= G_0 K_1$, $\TT^P_1= G_0 T^P_1$, and $\TT^C_1= G_0 T^C_1$. 
 Using the separable form of the pole term,
 \be
 T_1^P=\Gamma_1D_1\bar\Gamma_1,
 \ee
 where the diquark form factors are now antisymmetric, $P_{23} \Gamma_1 = -\Gamma_1$ and $ \bar \Gamma_1 P_{23}= -\bar \Gamma_1$, we obtain the equation for the Faddeev component
\be
\Phi_1= - \bar\Gamma_1P_{12} \left[1+G_0(K_1+T^C_1)P_{12}\right]^{-1} G_0 \Gamma_1D_1\Phi_1  \eqn{B10-P}
\ee
where
\be
\Phi_1=\bar\Gamma_1P_{12}\Psi_1.
\ee
Expanding the inverse term in \eq{B10-P} as
\be
\left[1+G_0(K_1+T^C_1)P_{12}\right]^{-1} = 1 - G_0(K_1+T^C_1)P_{12} + \dots,
\ee
leads to the final form of our unified equations for three identical quarks,
\be
\Phi_1= - \bar\Gamma_1P_{12} \left[1-G_0(K_1+T^C_1)P_{12}+ \dots\right] G_0 \Gamma_1D_1\Phi_1. \eqn{unified1}
\ee
Keeping only the first two terms of the series for the kernel, and making the further approximation, $T^C_1=0$, leads to the equation
\be
\Phi_1=\bar\Gamma_1P_{12} \left[1+K_1P_{12}\right] \Gamma_1d_1\Phi_1
\ee
which covers both approaches of Refs.\ \cite{Cahill:1988dx, Reinhardt:1989rw} and  Ref.\ \cite{Ebert:1996ec}.

\section{Discussion}
We have derived covariant equations that describe the bound state of the triquark in terms of quark and diquark degrees of freedom. These equations are illustrated in \fig{kvin}, with exact expressions given for distinguishable quarks in \eq{unified}, and for indistinguishable quarks in \eq{unified1}. An essential aspect of these equations is that they are exact in the approximation where only two-body forces are retained.
As a result, they are expected to encompass, and thus unify, all descriptions of triquarks that use quark and diquark degrees of freedom, and that assume two-body forces only.

It is worth noting that our procedure leading to \eq{3q-F+G}, and hence to \eq{unified} and \eq{unified1}, is similar to the one used by Alt, Grassbeger, and Sandhas (AGS) to reduce three-particle Faddeev equations to that of coupled two-particle equations  \cite{Alt:1967fx}; however, it differs from AGS in its details, and also in one essential way, namely, we have shown that the two-body matrix (in three-body space) $T_a$, can be decomposed into three mutually exclusive parts, as in  \eq{Tdecomp}, where the two-body kernel $K_a$ appears explicitly (AGS and related prior works, decomposed two-body t matrices into two parts, a separable one, and the rest). It is just this decomposition of $T_a$ into three-parts involving $K_a$, that has led to the unification of previous works, as outlined above.

This unification is demonstrated explicitly for two of the most prominent and longest-used approaches in the literature, namely  the one using the covariant quark+diquark Faddeev equations of Refs. \cite{Cahill:1988dx,Reinhardt:1989rw},  and the one using quasipotential quark-diquark equations of Ref.\ \cite{Ebert:1996ec}. In particular, the covariant quark+diquark Faddeev equations correspond to keeping just the first term of the kernel in our equations [the one-quark-exchange diagram in \fig{kvin}], and the quasipotential quark-diquark equations correspond to keeping just the second term of the kernel in our equations [the $qq$ rescattering diagram in \fig{kvin}]. It is noteworthy that our equations reveal that these two approaches, which have been pursued separately for more than 25 years in order to model not only bound states of baryons, but also various types of baryon form factors (electromagnetic, axial-vector, scalar, etc.), use equations with two, different,  non-overlapping, kernels. Our equations indicate that it is the sum of the first two terms in the kernel (at least) that should have been used instead. 
Although this is not an issue for cases where only one pair of quarks (out of three possible pairs) can form a diquark, in which case only the kernel of the quasipotential quark-diquark equations contributes \cite{Ebert:1996ec,Ebert:2005xj}, it may be a serious problem for other case, like that of three identical quarks where both kernels contribute and therefore should be summed \cite{Faustov:2021qqf}.

\begin{acknowledgments}
 A.N.K. was supported by the Shota Rustaveli National Science Foundation (Grant No. FR17-354).

\end{acknowledgments}

\bibliography{/Users/phbb/Physics/Papers/refn} 

\end{document}